\documentclass[conference]{IEEEtran}
\IEEEoverridecommandlockouts
\usepackage{cite}
\usepackage{amsmath,amssymb,amsfonts}
\usepackage{algorithmic}
\usepackage{graphicx}
\usepackage{textcomp}
\usepackage{adjustbox}
\usepackage{float}
\usepackage{xcolor}
\def\BibTeX{{\rm B\kern-.05em{\sc i\kern-.025em b}\kern-.08em
T\kern-.1667em\lower.7ex\hbox{E}\kern-.125emX}}

\usepackage{hyperref}
\hypersetup{
    colorlinks=true,
    linkcolor=black,
    filecolor=black,      
    urlcolor=black,
    citecolor=black
    }
    
\usepackage{microtype}
\usepackage[none]{hyphenat}

\usepackage{colortbl}
\newcolumntype{M}[1]{>{\centering\let\newline\\\arraybackslash\hspace{0pt}}m{#1}}

\usepackage{enumitem}

\makeatletter
\newcommand\BeraMonottfamily{%
  \def\fvm@Scale{0.85}
  \fontfamily{cmtt}\selectfont
}
\makeatother

\usepackage{listings}
\begin{document}

\title{Deep learning based auto-tuning for database management system}

\author{
\adjustbox{max width=\textwidth}{
\begin{tabular}{ccc}
 Karthick Gunasekaran &&  Kajal Tiwari \\ 
\textit{College of Information and Computer Sciences} && \textit{College of Information and Computer Sciences}  \\ 
\textit{University of Massachusetts Amherst} && \textit{University of Massachusetts Amherst}
\end{tabular}}
\\[5mm]
\adjustbox{max width=\textwidth}{
\begin{tabular}{c}
 Rachana Acharya \\ 
\textit{College of Information and Computer Sciences} \\ 
\textit{University of Massachusetts Amherst}
\end{tabular}}
}

\maketitle

\begin{abstract}
The management of database system configurations is a challenging task, as there are hundreds of configuration knobs that control every aspect of the system. This is complicated by the fact that these knobs are not standardized, independent, or universal, making it difficult to determine optimal settings. An automated approach to address this problem using supervised and unsupervised machine learning methods to select impactful knobs, map unseen workloads, and recommend knob settings was implemented in a new tool called OtterTune and is being evaluated on three DBMSs, with results demonstrating that it recommends configurations as good as or better than those generated by existing tools or a human expert.In this work, we extend an automated technique based on Ottertune \cite{Dana_Van} to reuse training data gathered from previous sessions to tune new DBMS deployments with the help of supervised and unsupervised machine learning methods to improve latency prediction. Our approach involves the expansion of the methods proposed in the original paper. We use GMM clustering to prune metrics and combine ensemble models, such as RandomForest, with non-linear models, like neural networks, for prediction modeling.
\end{abstract}


\section{Introduction}

\subsection{Problem Statement}
The main objective of our work is to train a machine learning (ML) model from data collected from previous tunings, and usethe models primarily for (1) pruning the redundant metrics, (2) mapping unseen database workloads to previous workloads from which we can transfer experience, and (3) improving latency prediction through workload mapping.
We have achieved this with the following steps: 

\begin{itemize}
\item \textit{Pruning Redundant Metrics:} Since we have hundreds of internal and external DBMS-specific metrics, we limit our search space to a small set, which impacts latency significantly. This step speeds up the entire process and ensures that the model will fit in memory.
\item  \textit{Workload Mapping:} Work load mapping is to match the target DBMS’s workload with the most similar workload in its repository based on performance measurements. This is done by finding the nearest neighbor with the help of Euclidean distance and calculating the "score" for each workload by taking
the average of these distances over all metrics. Reusing experience reduces time and resources and also helps the model make an educated decision in the case of fewer observations in the current workload.
\item  \textit{Latency Prediction:} The ultimate goal is to improve latency prediction through workload mapping. Prediction modeling will be done through the regression model-Guassian Process Regression), but we'll also be experimenting with neural networks. 
\end{itemize}

\section{Approach}

The Overview of the entire system can be seen from the \autoref{fig:archi} below. The process consisted of two important steps. In the first step, important metrics were pruned and in the next step automated tuning was carried out through workload mapping.   
\begin{figure}[H]
\centering
\includegraphics[width=\linewidth]{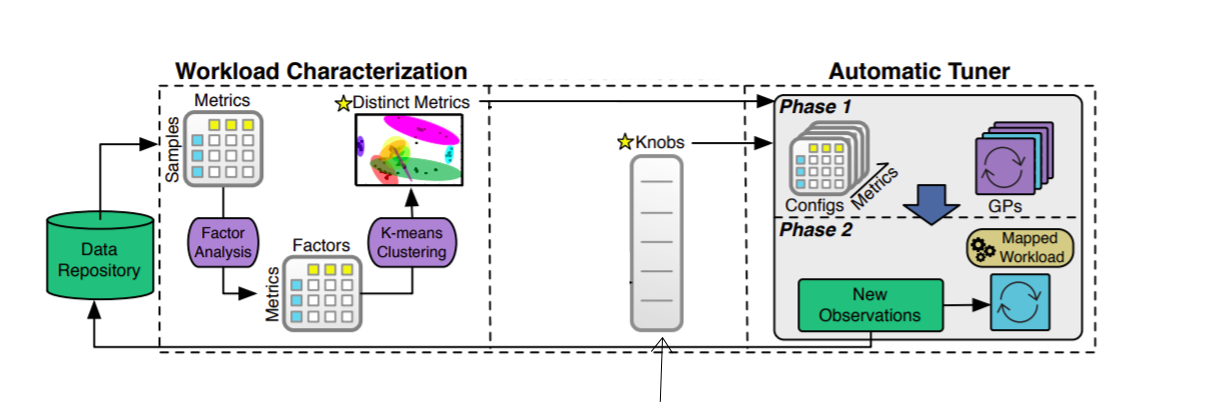}
\caption{Overview of the architecture}
\label{fig:archi}
\end{figure}

\subsection{Metrics pruning}

In this stage, pruning of redundant metrics was carried out. In order to capture the variability of system performance and differentiate different workloads, a small number of metrics with high variability are considered.  The system needs to capture a few metrics such that all the distinguishing characteristics of workloads are captured while the overall runtime of the machine learning algorithms reduces. The pruning step is very important because it helps improve the entire performance and speed of the system. Pruning is carried out in two steps. In the first step, factor analysis is carried out, followed by K-means clustering.

\subsubsection{Data Preprocessing}

We had to preprocess the data present in the offline workload and the online workloads B and C. Data pre-processing was done in the following steps:

\begin{itemize}
\item \textit{Duplicate Column Removal:} We dropped the columns or metrics with only a constant value across all workloads.
\item \textit{Conversion of boolean knobs:} We also did some other preprocessing steps, such as converting all boolean knob values to integers (0 and 1).
\item \textit{Division of files by workloads:}  We segregated all files on the workload ID; hence, for offline workloads, we had 58 workload files, for online workload C, we had 100 workload files; and for online workload B, we divided each workload into 5 rows for workload mapping and 1 row from each workload for validation.
\end{itemize}

\subsubsection{Factor Analysis}

Redundant attributes or metrics involve both the components that are strongly correlated and the metrics with different granularities. In this step, the strongly correlated metrics are removed. A dimensionality reduction technique called factor analysis \cite{sklearn_FA} is employed to transform the high dimensional data into low-dimensional data.

Taking a closer look at factor analysis, given a set of real-valued variables that contain arbitrary correlations, FA reduces these variables to a smaller set of factors that capture the correlation patterns of the original variables. 

Each factor is a linear combination of the original variables; the factor coefficients are similar to and can be interpreted in the same way as the coefficients in a linear regression. Furthermore, each factor has a unit variance and is uncorrelated with all other factors. This means that one can order the factors by how much of the variability in the original data they explain.

The FA algorithm takes as input a matrix $X$ whose rows correspond to metrics and whose columns correspond to knob configurations that we have tried. The entry $X_{ij}$ is the value of metric $i$
on configuration $j$. FA gives us a smaller matrix $U$: the rows of
$U$ corresponds to metrics, while the columns correspond to factors,
and the entry $U_{ij}$ is the coefficient of metric $i$ in factor $j$. We can
scatter-plot the metrics using elements of the $i^{th}$ row of $U$ as coordinates for metric $i$. Metrics $i$ and $j$ will be close together if they
have similar coefficients in $U$ - that is, if they tend to correlate
strongly in $X$. Removing redundant metrics now means removing
metrics that are too close to one another in our scatterplot.

\begin{figure}[htbp]
\centering
\includegraphics[width=\linewidth]{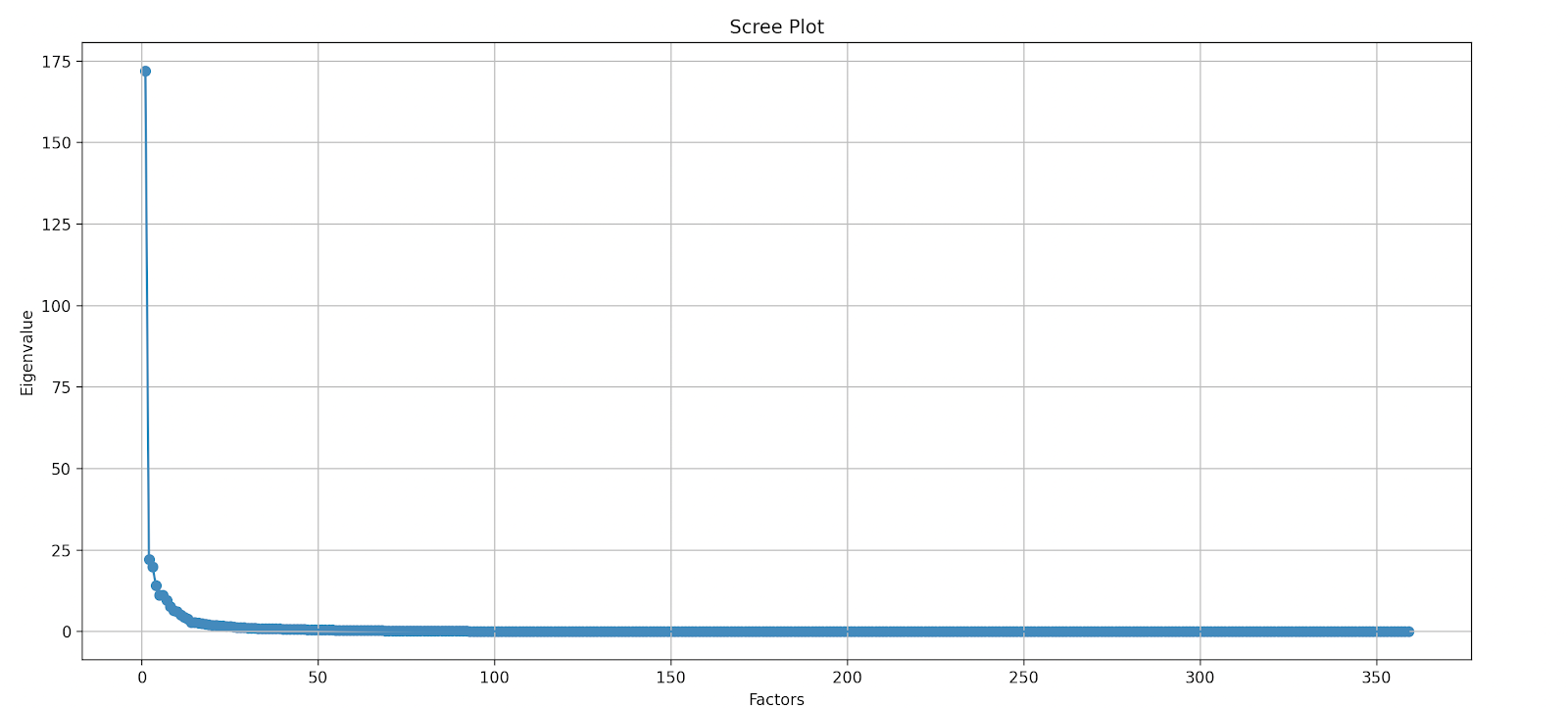}
\caption{Eigen values of every factor}
\label{fig:scree_plot}
\end{figure}

The data from offline workloads, which consist of 58 workloads, is used for this. Here, all the data related to the knobs is removed, and this truncated data is provided as input to the factor analysis module. Having a look at the eigen values of all the factors generated, only the initial 30 factors with eigen values greater than 1 are observed to be significant for the DBMS metric data. This means that most of the variability is captured by the first 30 factors. Therefore, only the most significant factors are taken into account.
\autoref{fig:scree_plot} shows a eigen values  where the factors are plotted against their respective eigen values. 

\subsubsection{K-means Clustering}

K-means clustering is used to identify and segregate metrics into meaningful groups. It works by clustering data to separate data into n groups of equal variance. The algorithm works towards minimizing the distance of the data points within the cluster.The input to the K-means model consists of a matrix comprising of our 30 factors from factor analysis.  A metric closest to each of the cluster centers can be picked up. Sklearn implementation of the K-means algorithm \cite{sklearn_kmeans} was made use of.

K-means doesn't provide with the optimal no of clusters. Initially K-means clustering is carried with different clusters numbers from 2 to 15.Then the optimal number of clusters are identified. Instead of doing manual interpretations of the clusters, silhouette analysis was carried out to study the separation distance between the clusters and  identify the optimal clusters. Silhouette values are decided by measuring closeness of the points in a cluster to its neighbouring clusters. The values range from -1 to +1. The value closer to 1 indicate the decision boundary is too far to the neighbouring clusters while value close to -1 indicate the clusters are too close to each other. So, Silhouette scores for KMeans models with different cluster configurations ranging from 2 to 15 were calculated. The Highest silhouette scores near +1 indicates best cluster configuration and its selected. 

Since the optimal no of clusters where identified as 8, the metrics are  segregated into 8 clusters and the centroids are computed for the same. One metric from each cluster which is the closest to the centroid was selected. This metric will represent the entire cluster. Once we find our target metrics, we remove all the other redundant metrics from the data and proceed to the next step.
List of pruned metrics by this process include,

{\scriptsize \begin{itemize}[itemsep=-9pt]

\item 
\begin{lstlisting} 
driver.jvm.pools.Code-Cache.committed.avg_inc
\end{lstlisting}

\item 
\begin{lstlisting}
driver.jvm.pools.Code-Cache.committed.avg_period
\end{lstlisting}	

\item 
\begin{lstlisting}
driver.BlockManager.memory.maxMem_MB.avg
\end{lstlisting}

\item 
\begin{lstlisting}
driver.BlockManager.memory.onHeapMemUsed_MB.avg
\end{lstlisting}

\item 
\begin{lstlisting}
driver.LiveListenerBus.queue.executorMgmt.size.avg
\end{lstlisting}

\item 
\begin{lstlisting}
executor.jvm.pools.PS-Old-Gen.cmt.avg
\end{lstlisting}

\item 
\begin{lstlisting}
worker_1.Disk_transfers_per_second.dm-0
\end{lstlisting}

\item 
\begin{lstlisting}
worker_1.Paging_and_Virtual_Memory.pgpg
\end{lstlisting}

\end{itemize}}

\subsection{Automated Tuning}

\subsubsection{Preprocessing}

All the pruned metrics other than the "latency" metric and the  knobs were normalized by removing the mean and scaling it to unit variance. Zero mean scaling is a very common practice before using the data in machine learning algorithms. This step is also essential to avoid overflow of values while calculating the mean squared error. 

\subsubsection{Gaussian Process Regression}

Gaussian process regression or GPR \cite{sklearn_GPR} a probabilitic based supervised learning approach was used to train the model. GPR requires prior to be specified. The prior in this case is the offline workloads which were seen before and they are fitted initially. The covariance of the prior are specified by the kernal choosen. Log marginal likelihood is maximized to optimize the hyperparameters during the fitting. Two different kernels were tried and the one which performed best was choosen. Radial basis function (RBF) performed better compared with others. The scaled knobs and metrics were used to fit the model. Sklearn implementation of the Gaussian process regression \cite{sklearn_GPR} as well Ottertuner implementation of the GPR was also used.

\subsubsection{Workload Mapping}

Ideally, the next component is the selection of knobs which have the highest impact on the system. This has already been done for us and the dataset provided to us already have the important knobs.

In workload mapping current target workload is mapped to previously gathered workloads. In our case, we mapped offline workload to both online workloads B and C by finding nearest neighbour with help of Euclidean distance and calculating 'score' for each workload by taking the average of these distances over all metrics.

The step by step process followed for Workload mapping is:

\begin{itemize}
\item  Calculate the Euclidean distance between the metrics vector for the target workload (workload B in 1st iteration) and the corresponding vector for
each workload  in the offline workload. 
This computation is done for each metric.
\item  Compute a 'score' for each workload i by taking the average of these distances over all metrics. The algorithm then chooses the workload with the lowest score as the one that is most similar to the target workload.
\item Augment the target workload to nearest source workload, in case of conflict of knob configuration between nearest source and  target workload, keep the configurations of target workload.
\item Train the models on 'Augmented workloads' and predict the latency on validation set created from online workload B.
\item Repeat the process with source workload as 'Augmented workload' and target workload as 'Workload C' and create final augmented workload for training the models and final prediction on test.csv.
\end{itemize}

\subsubsection{Latency Prediction}

Latency prediction is carried in two stages. In first stage, the 6th row of all the workloads in the online\_workloadB is used to predict the latency while trained on offline workloads appended with workload mapped data from online\_workloadB.The latency values are then estimated. In the next stage, the test.csv workloads are predicted on with the model trained on data appended from both workloadB and workloadC. This process adds more data for the GPR model to make accurate predictions.

\section{Extensions}

\subsection{GMM Clustering}

The original OtterTune paper \cite{Dana_Van} makes use of Kmeans for clustering metrics. But, there are certain disadvantages of using kmeans. It assumes that the clusters always have a spherical shape. Moreover, it takes only mean of of the clusters into account and not it's variance. Also, each cluster also has roughly equal number of observations. Additionally, it is also Very sensitive to outliers. To overcome these inadequacies, we replace the Kmeans clustering with GMM clustering \cite{sklearn_GMM}. 

GMM clustering overcomes these disadvantages as assuming each cluster as a different gaussian distribution and grouping data points belonging to a single distribution together. It therefore generates better clusters as it takes into account variance along with the mean of the clusters. One can think of mixture models as generalizing k-means clustering to incorporate information about the covariance structure of the data as well as the centers of the latent Gaussians. The GaussianMixture model used by us from sklearn \cite{sklearn_GMM} implements the expectation-maximization (EM) algorithm for fitting mixture-of-Gaussian models.  Expectation-maximization is a well-founded statistical algorithm to get around this problem by an iterative process. First one assumes random components (randomly centered on data points, learned from k-means, or even just normally distributed around the origin) and computes for each point a probability of being generated by each component of the model. Then, one tweaks the parameters to maximize the likelihood of the data given those assignments. Repeating this process is guaranteed to always converge to a local optimum. The GaussianMixture module comes with different options to constrain the covariance of the difference classes estimated: spherical, diagonal, tied or full covariance. The full covariance option is made us of by us, where each component has its own general covariance matrix, so the clusters may independently adopt any position and shape. 

The optimal number of clusters were found making use of the Silhoutte and BIC scores for GMM. 

The final set of metrics which were obtained were:

{\scriptsize \begin{itemize}[itemsep=-9pt]

\item 
\begin{lstlisting} 
driver.BlockManager.memory.onHeapMemUsed_MB.avg
\end{lstlisting}

\item 
\begin{lstlisting} 
driver.BlockManager.memory.remainingMem_MB.avg
\end{lstlisting}

\item 
\begin{lstlisting} 
driver.DAGScheduler.job.allJobs.nb_change
\end{lstlisting}

\item 
\begin{lstlisting} 
driver.DAGScheduler.stage.waitingStages.avg
\end{lstlisting}

\item 
\begin{lstlisting} 
driver.jvm.non-heap.usage.avg
\end{lstlisting}

\item 
\begin{lstlisting} 
driver.jvm.pools.Code-Cache.usage.avg
\end{lstlisting}

\end{itemize}}

\subsection{Random Forest algorithm}

Random forest algorithm was tried instead of Gaussian Process regression. One of the advantages with Random forest over Gaussian process regression is that it performs very well with high dimensional data. Also, the random forest is based on bagging algorithm and uses the emsemble technique where it creates many trees on the subset of the data and combines all the output. The overfitting is reduced through this technique, while GPR provides no such mechanism as the prior is fitted initially. Another reason to try random forest to make use of a  frequentist based approach rather than a probabilistic approach such GPR since frequentist based approach performs better when there is more data. Also due to their complexity, GPR require much more time to train than Random forest. Even though in this work there is comparatively less data being used when there is more data Random forest can be the right technique and might be preferable. Random forests algorithm \cite{sklearn_RF} was implemented with depth upto 50 levels and 200 estimators.

\subsection{Neural Network}

One major disadvantage with GPR used in OtterTune paper\cite{Dana_Van} is the fact that it doesn't perform well with higher dimensional data. So the no of metrics have to be pruned and used in ottertuner. To overcome this disadvantage Neural Networks is tried out as an alternative to see the performance over higher dimensions. An artificial neural network has the ability to learn and model non-linear and complex relationships which contain trainable parameters. Moreover, an artificial neural network’s outputs aren't limited entirely by inputs and results given to them initially by an expert system. Also, artificial neural networks have the ability to generalize their inputs. After learning from the initial inputs and their relationships, it can infer unseen relationships on unseen data as well, thus making the model generalize and predict on unseen data. 

A simple 2 layer neural network model is created with a fully connected hidden layer. The Keras library is made use for this purpose. No activation function is used for the output layer as we are interested in predicting numerical values directly without transformation. The efficient ADAM optimization algorithm is used and MAPE loss function is optimized. This metric is specifically chosen as this is the metric which will be used to compare all of our models. Adding another hidden layer was also tried out, but did not improve on the MAPE score. 

\section{Experiments}

\subsection{Evaluation}

In K-means clustering, the silhoutte analysis was carried to evaluate the optimal no of clusters in the data. 

In workload mapping stage, both Mean squared error and Mean average percentage error were tried for evaluation purposes. From the latency prediction results it was observed that the MSE evaluation criteria in workload mapping performed better.In latency prediction evaluation was carried out by using MAPE.

\subsection{Scaling and Hyperparameter tuning}

Experiments were carried with and without scaling. Two different scaling variations were performed and analysed. In first variation only the input for workload mapping and GPR models were tried. In second variation all the data except the latency column was scaled. In general, Scaling all the data except the latency column with zero mean and unit variance tend to improve the performance of the system. When unscaled data was used it was observed that the predictions were far apart from the ground truth in general.

Hyperparameter tuning was carried out for GPR.In case of GPR the noise levels were controlled by setting the alpha parameter to lower levels of values. From the \autoref{tab:gpralpha} it could be seen that the model performed well as the alpha values are low. Random forest trees were built for different depths and different estimators were also evaluated. The one with the best evaluation scores were choosen as hyperparameters.

\begin{table}[htbp]
\caption{GPR hyperparameter tuning}
\label{tab:gpralpha}
\centering
\adjustbox{max width=\linewidth}{\fontsize{8}{14}\selectfont \it
\begin{tabular}{M{4cm}M{4cm}} \hline \hline 
\textbf{Alpha} & \textbf{MAPE} \\ \hline 
1e+8 &  168.63 \\  
1e+7 &  142.01 \\  
1e+5 &  123.67 \\  
1e+3 &  101.29 \\  
1e+1 & 98.45 \\  
1e-1 & 69.61\\ \hline\hline 
\end{tabular}}
\end{table}

We also experimented with adding an additional hidden layer for the Neural Network. Though, this did not improve on the MAPE scores.

\section{Results}

\autoref{tab:structured} shows results obtained from different models evaluated on the testset from online workload B. From the table we can see that replacing K-means by GMM clustering improved the performance of the model by slight extent by both metrics MAPE and MSE. However the performance of both random forest and the neural networks are not as good as GPR. This is mainly due to the lesser amount data involved in the work. This can be clearly seen with the MSE values seen in the \autoref{tab:structured}. Neural networks have a higher tendency to overfit when given lesser data which could be verified with the higher MSE value.Our experiments in general show that results slightly improved by using GMM clustering.

\begin{table}[htbp]
\caption{Results summary}
\label{tab:structured}
\centering
\adjustbox{max width=\linewidth}{\fontsize{9}{14}\selectfont \it
\begin{tabular}{M{3.25cm}M{2.5cm}M{2.5cm}} \hline \hline
\textbf{Type} & \textbf{MAPE} & \textbf{MSE} \\ \hline
Baseline &  69.61 & 2118 \\ 
EM clustering & 67.85 & 2329 \\ 
Random Forest & 78.98 & 3817 \\  
Neural Network &  77.26 & 13426 \\ \hline\hline 
\end{tabular}}
\end{table}

\begin{figure}[htbp]
\centering
\begin{tabular}{c}
\includegraphics[width=0.9\linewidth,height=0.18\paperheight]{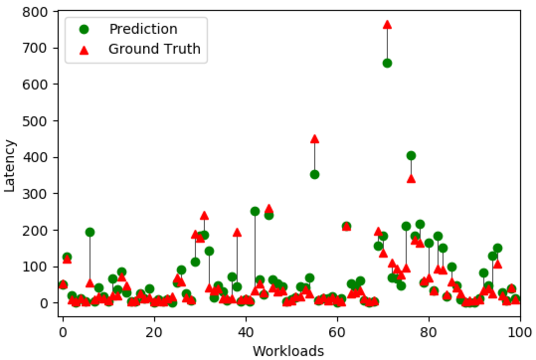} \\
(a) \\ [0.5mm]
\includegraphics[width=0.9\linewidth,height=0.18\paperheight]{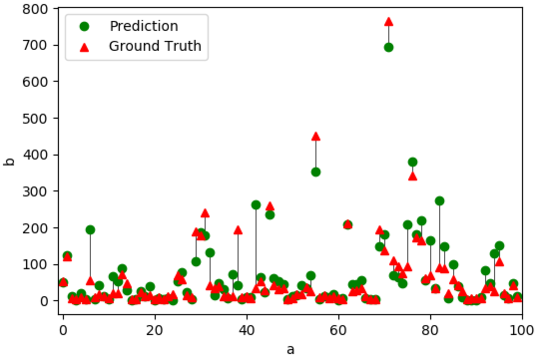} \\ 
(b) \\ [0.5mm]
\includegraphics[width=0.9\linewidth,height=0.18\paperheight]{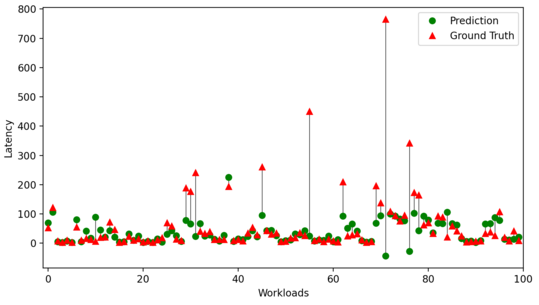} \\ 
(c) \\ [0.5mm]
\includegraphics[width=0.9\linewidth,height=0.18\paperheight]{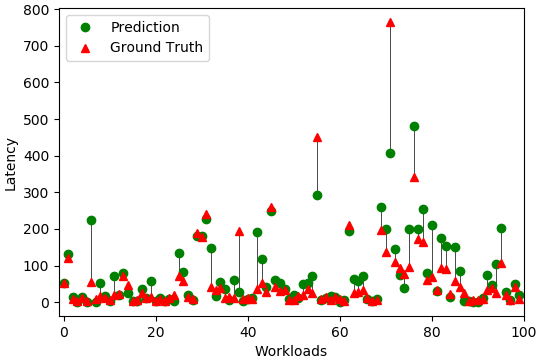} \\ 
(d) 
\end{tabular}
\caption{Predictions and Ground truth latency prediction on test set of online Workload B (a) Baseline (b) EM clustering  (c) Neural networks (d) Random forest}
\label{fig:baseline_emclustering_NN-2_random}
\end{figure}

\autoref{fig:baseline_emclustering_NN-2_random} shows the ground truth and model predictions for latency on online Workload B for both Baseline, EM clustering and neural network. The plot shows that the predictions made using the metrics from EM-clustering algorithms was much more closer to the ground truth when compared with the baseline approach. However the predictions are very far apart in case of neural networks.

\section{Conclusion}

Automatic DBMS tuning remains an active area of research and we presented an automatic approach which leverages past experience and collects new information to tune DBMS configurations. During experiments, we were able to achieve MAPE of 69\% using baseline implementation which utilizes FA,Kmeans for metric pruning and GPR for prediction modelling. 

Our experiments also show that by replacing K-means clustering with EM-clustering, we were able to reduce MAPE to 67\% which suggest that GMM clustering could be explored further as an alternative to K-means clustering. Although the dataset we're working on is very small to make a definitive statement that EM clustering is better than KMean clustering, we propose that EM clustering should be explored further as K-means alternative as EM clustering works well with non-linear geometric distributions and doesn't assumes shape of clusters to be spherical as in the case of K-means.

We also experimented with ensemble methods like Random Forest with resulting MAPE of 78\% and non-linear models like Neural network with resulting MAPE of 77\%. However in both of these cases, our baseline model with GPR performs better than neural and RandomForest model. But an good future work will be to collect more data and use it the same approaches.

\end{document}